\documentclass[aps,twocolumn,showpacs]{revtex4}%
\usepackage{amsfonts}
\usepackage{amsmath}
\usepackage{amssymb}
\usepackage{graphicx}%
\begin{document}
\title{Experimental estimates of dephasing time in molecular magnets.}
\author{Amit Keren and Oren Shafir}
\affiliation{Physics Department, Technion-Israel Institute of Technology, Haifa 32000,
Israel }
\author{Efrat Shimshoni}
\affiliation{Department of Mathematics-Physics, University of Haifa at Oranim, Tivon 36006, Israel}
\author{Val\'{e}rie~Marvaud, Anne Bachschmidt, and J\'{e}r\^{o}me Long}
\affiliation{Laboratoire de Chimie Inorganique et Mat\'{e}riaux Mol\'{e}culaires (UMR CNRS
7071) Universit\'{e} Pierre et Marie Curie, 75252 Paris, France}
\pacs{75.50.Xx, 76.75.+i}

\begin{abstract}
Muon spin relaxation measurements in isotropic molecular magnets (MM) with
spin value $S$ ranging from $7/2$ to $27/2$ are used to determine the
magnitude and origin of dephasing time $\tau_{\phi}$ of molecular magnets. It
is found that $\tau_{\phi}\sim10$~nsec with no $S$ or ligand dependence. This
indicates a nuclear origin for the stochastic field. Since $\tau_{\phi}$ is a
property of the environment, we argue that it is a number common to similar
types of MM. Therefore, $\tau_{\phi}$ is shorter than the Zener and tunneling
times of anisotropic MM such as Fe$_{8}$ or Mn$_{4}$ for standard laboratory
sweep rates. Our findings call for a stochastic Landau-Zener theory in this
particular case.

\end{abstract}
\date{June 22, 2007}

\maketitle

\begin{figure}[h]
\begin{center}
\includegraphics[
natheight=6.371100in,
natwidth=7.721900in,
height=2.9897in,
width=3.6175in
]%
{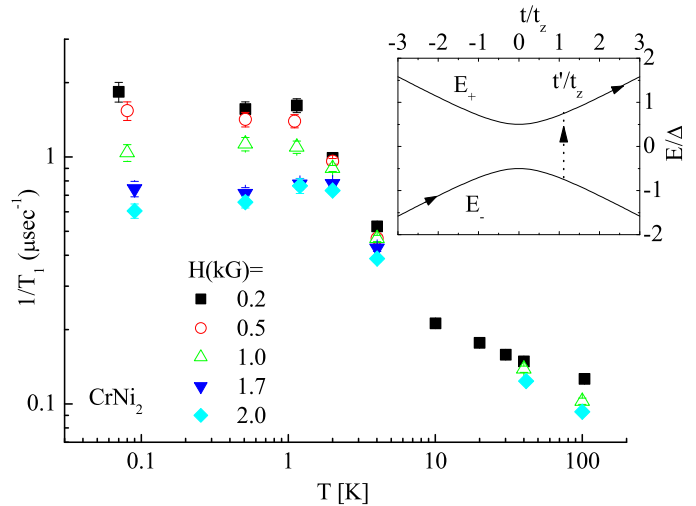}%
\caption{(color online). The muon spin lattice relaxation rate $1/T_{1}$ as a
function of temperature and field in the CrNi$_{2}$ ($S=7/2$) high spin
molecular magnet. Inset: the solid lines show the instantaneous energy levels
as a function of time in the Landau-Zener problem. Dashed line is a schematic
representation of a path the spin can take when tunneling from the low energy
state to the high energy state at time $t^{\prime}/t_{Z}$.}%
\label{crni2t1vst}%
\end{center}
\end{figure}

Quantum tunneling of the magnetization in anisotropic molecular magnets (MM)
with high spin value is a fascinating subject which contrasts clean and
accurate experimental data with sophisticated theoretical models \cite{Intro}.
At the heart of these models stands the Landau \cite{LandauPZS32} and Zener
\cite{ZenerPPSL32} (LZ) derivation of quantum tunneling between levels, which
at resonance have a tunnel splitting $\Delta$, but are brought into and out
off resonance by a time-dependent field. This model can be described by the
Hamiltonian $\mathcal{H}_{0}=\beta tS_{z}+\Delta S_{x}$, where $\mathbf{S}$ is
the electronic spin operator and $\beta$ is proportional to the external field
sweep rate $dH/dt$. The LZ theory predicts the transition amplitude $C_{LZ}$
that a spin prepared at time $t=-\infty$ in the low energy state $\left\vert
+\right\rangle $ will be in the high energy state at $t=\infty$ which is again
$\left\vert +\right\rangle $, namely, $C_{LZ}=\left\langle +\right\vert
U\left\vert +\right\rangle $ where $U$ is the time propagator operator. The
calculation of this amplitude has a path integral representation as
demonstrated graphically in the inset of Fig.~\ref{crni2t1vst}
\cite{KayanumaJPSJ84,ShimshoniPRB93,SinitisynPRB03}. In this inset, the solid
lines show the instantaneous energies $E_{\pm}=\pm\frac{1}{2}\sqrt{\Delta
^{2}+\beta^{2}t^{2}}$, and a single path is associated with a transition from
the lower energy state to the upper energy state, which occurs at a specific
time $t^{\prime}$.

However, there is consensus among researchers that the tunneling in MM is
incoherent due to interactions of the spin with a stochastic field
$\mathbf{B}(t)$ which is produced by nuclear moments
\cite{SinitisynPRB03,ProStamp,StampPRB04,WernsdorferEPL00,VillainEJP05}, and
that the dephasing time of the quantum states must be taken into account. The
dephasing time $\tau_{\phi}$ is defined using the corelator of the stochastic
field
\begin{equation}
\left\langle \mathbf{B}(t)\mathbf{B}(0)\right\rangle =\left\langle
\mathbf{B}^{2}\right\rangle \exp(-t/\tau_{c}), \label{FieldCorrelation}%
\end{equation}
as%
\begin{equation}
\frac{1}{\tau_{\phi}}=\frac{\left\langle B^{2}\right\rangle \tau_{c}}%
{\hbar^{2}}. \label{tauphi}%
\end{equation}
When the dephasing time is very long, the transition probability $P_{LZ}$ is
given by the absolute value square, of the sum of the transition amplitudes,
for different paths. This yields the famous expression
\begin{equation}
P_{LZ}=1-\exp\left(  -\frac{\pi\Delta^{2}}{2\hbar\beta}\right)  \label{LZ}%
\end{equation}
of flipping states \cite{ShimshoniBig}. In contrast, if the dephasing time is
very short, the interference between paths should be destroyed and transition
probability should become a sum of instantaneous transition probabilities.

Therefore, there are four important time scales in the LZ problem: I) the
tunneling time $t_{T}=\hbar/\Delta$ which is set by the tunnel spliting, II)
the Zener time $t_{z}=\Delta/\beta$, which is the time segment around $t=0$
where tunneling can occur during a field sweep in the adiabatic case
($t_{T}\ll t_{z}$), III) the correlation time $\tau_{c}$, and IV) the
dephasing time $\tau_{\phi}$ over which different paths interfere coherently.
Determining these time scales even roughly could help select the theory for
the analysis of magnetization jump experiments. Moreover, theories are
available only for particular orders of time scales, which might not be the
realistic ones.

The theories addressing the stochastic LZ problem can be divided into two
groups according to the type of stochastic field they use: Ising type with
coupling $B_{z}(t)S_{z}$, or Heisenberg type with an $\mathbf{B}%
(t)\mathbf{\cdot S}$ term. In the Ising case Kayanuma \cite{KayanumaJPSJ84}
found modifications to the LZ formula for the order of time scales $\tau
_{c}\ll\tau_{\phi}\ll(t_{z}t_{T})^{1/2}$ and $\tau_{c}\ll$Max$\left[
t_{z},t_{z}t_{T}/\tau_{\phi}\right]  $. In this case the transition
probability is give by $P=\left[  1-\exp\left(  -\pi t_{z}/t_{T}\right)
\right]  /2$. Therefore, when the transition is sudden ($t_{T}\gg t_{z}$) then
$P=\pi\Delta^{2}/(2\hbar\beta)$ as in Eq.~\ref{LZ} at the same limit.
Sinitsyn, Prokof'ev, and Bobrovitski \cite{SinitisynPRB03} extended this work
using macroscopic spin bath description of $B_{z}(t)$ and showed that
Kayanuma's sudden result is correct if and only if $t_{T}\gg t_{z}$. In the
Heisenberg case, Shimshoni and Stern found corrections to the LZ formula in
all orders of time scales they examined. Here we mention just the interesting
case of $\tau_{c}\ll\tau_{\phi}$ and $t_{T}\ll\tau_{\phi}\ll t_{z}$where they
found that $P\simeq1-(\tau_{\phi}/t_{Z})\left[  \exp(2t_{Z}/\tau_{\phi}%
)P_{LZ}+(t_{T}/\tau_{\phi})^{2}\right]  $ \cite{ShimshoniPRB93}. More
theoretical work can be found in Ref.~\cite{MoreTheory}. The consensus seems
to be that when the field sweep is adiabatic $t_{z}\gg t_{T}$, the stochastic
field modifies the LZ formula, and that in the sudden limit $t_{T}\gg t_{z}$
of the Ising case the dephasing time $\tau_{\phi}$ has no impact on the
tunneling probability. However, as far as we know there is no theory for the
Heisenberg coupling when $\tau_{\phi}$ is the shortest time scale in the problem.

Despite the importance of $\tau_{\phi}$ determination in the LZ problem, today
there is no experimental estimate of this time in the problem of magnetic
quantum tunneling. The purpose of the present work is to provide such an
estimate. We do so by measuring the dephasing times of isotropic molecular
magnets ($\Delta=0$) with different spin value and ligands, and project the
result to anisotropic MM such as Fe$_{8}$ or Mn$_{4}$. This allows us to set
the order of $t_{z}$, $t_{T}$, and $\tau_{\phi}$. Our major finding is that
$\tau_{\phi}$ is the shortest time scale in the problem. Since nuclear dipolar
coupling to the molecular spins involves all directions, we conclude that
there is no relevant theory for the LZ problem in MM with stochastic field.

In addition to the contribution to the problem of magnetic quantum tunneling,
our experiment has its own merit. It is the first examination of magnetic
fluctuation as a function of the spin value $S$. As such it provides a new
look at the interaction between spins and the lattice in the quantum
(temperature independent) regime.

We determine the dephasing times of isotropic molecules by performing muon
spin relaxation measurements on eight different MM with $\Delta\simeq0$ and
spin value ranging from $S=7/2$ to $S=27/2$. The major assumption here is that
$\tau_{\phi}$ is a property of the environment and not of the molecule (see
Eq.~\ref{tauphi}). Therefore, if we determine $\tau_{\phi}$ for one type of
molecule, and if a different molecule has the same environment, it will have
the same $\tau_{\phi}$. This assumption received experimental support recently
in the work of Ardavan \emph{et al.}. They showed using ESR that two different
molecules, one with zero field splitting and the other without it, have the
same electronic $T_{2}$ \cite{ArdavanPRL07}. However, it also has advocates.
Stamp, Tupitsyn, and Morello argue that the molecular electronic spin impacts
the nuclear spin dynamic and therefore the dephasing time should depend on
$\Delta$ so that $\tau_{\phi}\propto\Delta$ \cite{StampPRB04}. Encouraged by
the experimental finding we continue the presentation using our assumption.

What allows us to extract the dephasing time is the fact that our muons are
coupled to the electronic spins of isotropic MM that experience only the
stochastic and external fields. Therefore, the number of parameters needed to
be determined in our experiment is minimal, and there is no need to know
\textit{a priori} the order of time scales. The leading terms for such an
Hamiltonian are%
\begin{equation}
\mathcal{H}=-2\mu_{B}\left[  \mathbf{H}+\mathbf{B}(t)\right]  \mathbf{S}%
+\hbar\nolinebreak^{\mu}\gamma\left[  \mathbf{H}+\mathbf{SA}\right]
\mathbf{I} \label{MuonHamilt}%
\end{equation}
where $\mathbf{I}$ is the muon spin, $\mathbf{H}$ is the external field,
$^{\mu}\gamma=851.62$ ~MHz/T is the muon gyromagnetic ratio, $\mu_{B}$ is the
Bohr magneton, and $\mathbf{A}$ is a coupling matrix. We ignore the
$\mathbf{B}(t)\mathbf{I}$ term since the field experienced by the muon from
the molecular spins is greater than this term. Due to the fluctuating field
$\mathbf{B}$, $\mathbf{S}$ will vary in time. The simplest assumption that one
can make is that the correlation function $\left\langle \left\{
\mathbf{S}(t),\mathbf{S}(0)\right\}  \right\rangle $, where $\left\{
{}\right\}  $ stands for anticommutator, decays exponentially. The decay rate
is determined by the dynamic properties of $\mathbf{B}(t)$ which is produced
by the environment of the molecules. Therefore, we expect%
\begin{equation}
\left\{  \mathbf{S}(t),\mathbf{S}(0)\right\}  =2S^{2}\exp(-t/\tau_{\phi})
\label{SpinAuto}%
\end{equation}
with $\tau_{\phi}$ set by Eq.~\ref{tauphi}. It is possible that $\tau_{\phi}$
will be $H$ dependent but we will show experimentally that this is not the
case for $H\leq2$~kG.

We investigated CrCu$_{4}$ ($S=7/2$), CrNi$_{2}$ ($S=7/2$), CrNi$_{2}$Mn$_{4}$
($S=13/2$), CrNi$_{2}$Ni$_{4}$ ($S=15/2$) and CrNiMn$_{5}$ ($S=20/2$). To
this, we added data from a previous study of CrCu$_{6}$ ($S=9/2$), CrNi$_{6}$
($S=15/2$), and CrMn$_{6}$ ($S=27/2$) by Salman \textit{et al.}
\cite{SalmanPRB02}. These compounds, based on polycyanometalated precursors,
are prepared following a step-by-step synthetic strategy. The key idea is to
use polydentate amine ligands in order to avoid polymerization and get
discrete entities with well-defined spin and anisotropy \cite{VM1,VM2}. Most
of the compounds are fully described in the literature \cite{VM3,VM4}. They
may be divided into two groups: i) isotropic high spin molecules (CrCu$_{6}$,
CrNi$_{6}$, and CrMn$_{6}$ ), ii) nearly isotropic molecules with no
detectable energy gap or small one $\sim$1~K (CrCu$_{4}$, CrNi$_{2}$,
CrNiMn$_{5}$, CrNi$_{2}$Mn$_{4}$, CrNi$_{2}$Ni$_{4}$).

In our $\mu$SR-$T_{1}$ experiments we measure the polarization $P(t,H)$ of a
muon spin implanted in the sample, as a function of time $t$ and magnetic
field $H$, when the field is applied in the direction of the initial muon
polarization. These experiments were performed at both ISIS and PSI,
exploiting the long time window in the first facility for the slow relaxation
of the low $S$ molecules, and the high time resolution in the second facility
for the fast relaxation of the high $S$ molecules. Typical raw $\mu$SR data
are presented in Ref.~\cite{SalmanPRB02}. The data for all samples are fitted
to $P(H,t)=\exp(-\sqrt{t/T_{1}})+Bg$, where $Bg$ is a field and temperature
independent background. This root exponential behavior is a consequence of the
many different muon sites in the sample.%



In Fig.~\ref{crni2t1vst} we depict the temperature dependence of $1/T_{1}$. As
the temperature is lowered, the relaxation increases due to slowing down of
the spin fluctuation as a result of the interactions between spins in the
molecules. However, once the MM is formed, the spin dynamics is nearly
temperature independent down to the milikelvin regime. All molecules show the
same behavior. More raw $T_{1}$ data are presented in Ref.~\cite{SalmanPRB02}.

\begin{figure}
[ptb]
\begin{center}
\includegraphics[
natheight=7.971800in,
natwidth=6.384000in,
height=3.2923in,
width=2.642in
]%
{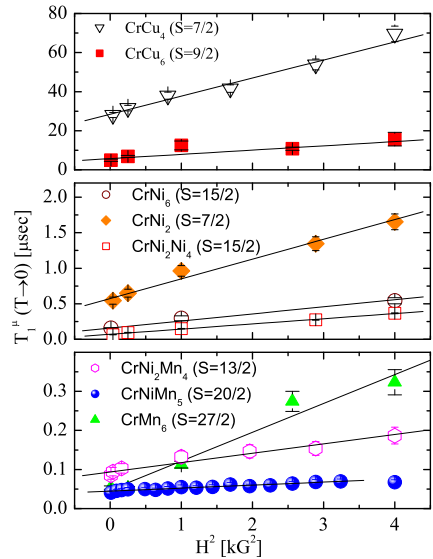}%
\caption{(color online). Muon $T_{1}$ at 100~mK versus field squared for all
molecules including three from Ref.~\cite{SalmanPRB02}. The solid lines are
linear fits.}%
\label{t1vsh2}%
\end{center}
\end{figure}

In Fig.~\ref{t1vsh2} we depict $T_{1}$ as a function of $H^{2}$ for all the
molecules measured to date. There is a large variation in the scale of $T_{1}$
between the different molecules. A linear dependence of the form
$T_{1}=m+nH^{2}$ is found in all cases, as demonstrated by the fitted solid
line. This is in agreement with Ref. \cite{SalmanPRB02}. The difference
between molecules is in the slope $n$ and crossing of the line $m$. The
dephasing time could be extracted from the standard theory of $T_{1}$
relaxation where
\begin{equation}
\frac{1}{T_{1}}=\frac{2A^{2}\tau_{\phi}}{1+(^{\mu}\gamma H\tau_{\phi})^{2}}.
\label{InvT1}%
\end{equation}
Although this expression is a result of perturbation expansion where
$\mathbf{H}$ provides the quantization axis, it was demonstrated by numerical
methods to be a good approximation even for $H\rightarrow0$ \cite{KerenPRB94}.
Here we assumed for simplicity that $\mathbf{A}$ is diagonal and isotropic,
but this assumption has no significance for our conclusions. $\tau_{\phi}$ is
obtained from
\begin{equation}
\tau_{\phi}=\left(  \frac{n}{m^{\mu}\gamma^{2}}\right)  ^{1/2}
\label{tauphifromfit}%
\end{equation}
for each molecule at the lowest temperature.%

\begin{figure}[t]
\begin{center}
\includegraphics[
natheight=6.398700in,
natwidth=8.015100in,
height=2.9888in,
width=3.736in
]%
{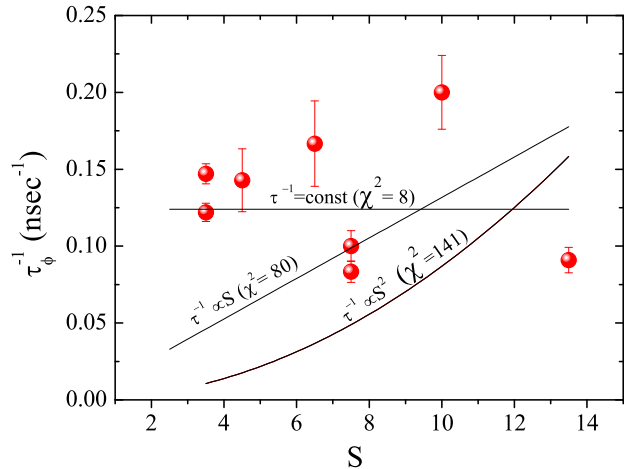}%
\caption{(color online). Dephasing rate $\tau_{\phi}^{-1}$ extracted from the
muon relaxation data as a function of spin value $S$ of the various magnetic
molecules. The solid lines are fits to power laws as indicated in the figure.
$\chi$ represent the quality of the fit.}%
\label{globalnu}%
\end{center}
\end{figure}

The main experimental finding of this work is presented in Fig.~\ref{globalnu}
where $\tau_{\phi}^{-1}$ is plotted as a function of $S$ for all the
molecules. This plot shows that within experimental errors $\tau_{\phi}$ is
weakly dependent on the type of molecule used, despite the large variations in
$T_{1}$. In particular, $\tau_{\phi}$ is weakly dependent on $S$ or the
ligand. To emphasize this conclusion we fit the data to three different power
laws: $\tau_{\phi}^{-1}\propto$ constant, $S$, and $S^{2}$. The quality of the
fit expressed as the value of the reduced $\chi^{2}$ is shown on the graph.
The $\tau_{\phi}^{-1}=const$ gives an order of magnitude better fit than the
other power laws. It is also interesting to compare our finding of $\tau
_{\phi}\sim10$~nsec to other experiments. In the deuterated molecules Cr$_{7}%
$Mi and Cr$_{7}$Mn the ESR $T_{2}$ (interpreted here as $\tau_{\phi}$) is
$3$~$\mu$sec \cite{ArdavanPRL07}. Had the samples were not deuterated,
$\tau_{\phi}$ would have been $80$~nsec due to the gyromagnetic ratio between
protons and deuterium. In the V$_{15}$ molecule $\tau_{\phi}\sim2$~nsec
\cite{V15}.%

It is highly significant that $\tau_{\phi}$ is nearly spin- and
ligand-independent. Since $\tau_{\phi}$ is determined by the environment in
which the molecules are embedded, its $S$-independence means that coupling to
other molecules or to phonons is not responsible for $\tau_{\phi}$. In both
these mechanisms the field $\mathbf{B}$ experienced by a given molecule
depends on $S$, and according to Eq.~\ref{tauphi} we would expect an
$S$-dependent $\tau_{\phi}$. We therefore conclude that at $T\rightarrow0$ the
stochastic field $\mathbf{B}(t)$ responsible for the MM spin motion emanates
from nuclear moments, most likely protons. Since there are many protons in the
ligands, the variations between ligands do not have a big impact on
$\tau_{\phi}$. According to Eq.~\ref{tauphi}, $\tau_{\phi}$ on the order of
$10$~nsec could be generated by a field $B\sim$ $1$ to $0.01$~G, which for
$S=10$ is equivalent to $200$ to $0.2$ MHz, fluctuating at a rate of
$1/\tau_{c}\sim4$ to $4\times10^{-4}$~$\mu$sec$^{-1}$, respectively. These
values are typical for nuclei. In Fe$_{8}$ Morello \textit{et al.} found
nuclear $1/T_{2}$ on the order of $10^{-4}$~$\mu$sec$^{-1}$
\cite{MorelloPRL04}.

As we argued before, the dephasing time should be typical of high spin
magnetic molecules made of transition metal ions embedded in a sea of protons.
Indeed, the eight isotropic molecules reported here are different but have
similar $\tau_{\phi}$. We have no \emph{experimental} reason to believe that
$\tau_{\phi}$ will be substantially different in Fe$_{8}$ or Mn$_{4}$ where
$\Delta$ was measured. In both cases $\Delta\sim10^{-7}$~K for the $-S$ to $S$
transitions \cite{WorensdoferScience99}. The tunneling time $t_{T}%
=\hbar/\Delta\sim7.6\times10^{-5}$~sec. This tunneling time is longer than the
dephasing time $\tau_{\phi}\sim10^{-8}$~sec in our, and other
\cite{ArdavanPRL07,V15} molecules. Moreover if, for example, $\beta
=0.001$~K/sec, then $t_{z}=\Delta/\beta\sim1\times10^{-4}$~sec (for the same
transition). This implies the order of time scales $t_{Z}\sim t_{T}\gg
\tau_{\phi}$, a regime which corresponds to a strong dephasing. For
$\beta=0.1$~K/sec we have $t_{T}\gg t_{Z}\gg\tau_{\phi}$. As we mentioned
before, the impact of the Heisenberg type stochastic fluctuations in this
order of time scales on transition probabilities is not known theoretically.

To summarize, we have measured spin correlations in isotropic molecular
magnets on a wide range of $S$ values. We found that the correlation time is
nearly $S$- and ligand-independent and on the order of $10$~nsec. We use this
time as an estimate of dephasing times in non isotropic molecules such as
Fe$_{8}$ and Mn$_{4}$ where tunneling occurs. Our findings call for a
theoretical development of the LZ problem with stochastic field fluctuations
coupled to all components of the spin $\mathbf{S}$ operator, where $\tau
_{\phi}$ is the shortest time scale in the problem.

We are indebted to W. Wernsdorfer for helpfull discussions. We also
acknowledge financial support from the Russell Berrie Nanotechnology Institute
in the Technion, the Israeli ministry of science, and the European Commission
under the 6th Framework Programme. We are also grateful for the ISIS and PSI
facilities for high quality muon beams and technical support.

\end{document}